\documentclass[twocolumn,showpacs,amsmath,amssymb,prl,superscriptaddress,floatfix]{revtex4-1}
\usepackage{graphicx}% Include figure files
\usepackage{bm}% bold math
\usepackage{amsmath,amsfonts}

\begin{document}

\title{Many-body spin interactions and the ground state of a dense Rydberg lattice gas}

\author{Igor Lesanovsky}

\affiliation{School of Physics and Astronomy, University of
Nottingham, Nottingham, NG7 2RD, UK}


\date{\today}

\begin{abstract}
We study a one-dimensional atomic lattice gas in which Rydberg atoms are excited by a laser and whose external dynamics is frozen. We identify a parameter regime in which the Hamiltonian is well-approximated by a spin Hamiltonian with quasi-local many-body interactions which possesses an exact analytic ground state solution. This state is a superposition of all states of the system that are compatible with an interaction induced constraint weighted by a fugacity. We perform a detailed analysis of this state which exhibits a cross-over between a paramagnetic phase with short-ranged correlations and a crystal. This study also leads us to a class of spin models with many-body interactions that permit an analytic ground state solution.
\end{abstract}
%\pacs{}
\maketitle
Strong interactions competing with non-commuting single particle terms in a many-body quantum Hamiltonian often lead to non-classical ground states. Only in exceptional cases analytic or approximately analytic results can be found. Paradigm examples are the one-dimensional $xy$-model in a transverse field \cite{Sachdev99} or the celebrated Affleck-Kennedy-Lieb-Tasaki spin model \cite{Affleck88}, both of which have proven indispensable for the understanding of many-body phenomena in magnetic compounds and valence bond solids. Finding models of experimentally realizable many-body Hamiltonians with exact solutions is hence of fundamental interest.

Models of many-body quantum systems with origin in condensed matter physics are currently implemented and studied in ultra cold atomic systems with great success \cite{Bloch08}. Most experiments so far are carried out with ground state atoms, but very recent efforts exploit the unique properties of atoms excited to Rydberg states. These states offer strong and long-ranged interatomic interactions in conjunction with an extraordinarily long lifetime \cite{Gallagher84,Saffman10-2}. This enabled remarkable experiments which studied the coherence properties in strongly interacting Rydberg gases \cite{Heidemann07,Reetz-Lamour08} and eventually demonstrated the feasibility to process quantum information with Rydberg atoms \cite{Urban08,Gaetan08}. Rydberg gases are moreover an almost ideal experimental implementation of interacting spin systems such as the aforementioned $xy$-model \cite{Olmos09-3}. This stimulated a plethora of theoretical studies investigating the real time evolution \cite{Weimer08,Lesanovsky10} as well as ground states of these spin models \cite{Pohl10,Schachenmayer10,Weimer10-2}. The latter, predominantly numerical work, revealed a variety of interesting quantum phases and studied the creation \cite{Pohl10,Schachenmayer10} as well as the melting dynamics \cite{Weimer10-2} of dynamically created crystals.

In this work we provide an analytic study of the non-trivial entangled many-body ground state of a strongly interacting one-dimensional Rydberg gas. The strong interaction among excited atoms gives rise to an effective Hamiltonian with a quasi-local three-body interaction that effectuates a set of non-commuting local constraints. For certain values of the experimental parameters this Hamiltonian is accurately approximated by a spin Hamiltonian which has an exact analytical ground state solution. We show that this is a consequence of the fact that the Hamiltonian possesses a manifold of approximate Rokhsar-Kivelson points \cite{Rokhsar88} where is assumes a so-called Stochastic Matrix Form \cite{Castelnovo05}. The ground state is a coherent superposition of all states compatible with all the local constraints weighted by an effective fugacity. We analytically explore the properties of this state which shows a cross-over between a paramagnetic phase with short-ranged correlations and a crystalline ordered state. Our study highlights a new perspective for creating and studying non-classical and entangled states with ultracold Rydberg gases. It also leads to a class of spin models with many-body interaction whose non-trivial ground state solution can be found analytically.

Our system consists of a deep one-dimensional lattice with $L$ sites evenly spaced at a distance $a$. For convenience we consider periodic boundary conditions, but this is no necessary requirement. Each site is occupied by a single atom which we treat within the two-level approximation where every atom forms a (pseudo)spin $1/2$ particle. The atomic ground state $\left|g\right>\equiv\left|\downarrow\right>$ is coupled to a Rydberg state $\left|r\right>\equiv\left|\uparrow\right>$ by a laser with Rabi frequency $\Omega$ and detuning $\Delta$. Atoms in Rydberg states interact via a power law interaction with (inverse) exponent $\gamma > 0$. The Hamiltonian of this system is (within the rotating wave approximation for the atom-laser coupling) given by
\begin{eqnarray}
  H_0=\Omega \sum_k^L \sigma_x^k + \Delta \sum_k^L n_k + V \sum_{m>k} \frac{n_m n_k}{|k-m|^\gamma}. \label{eq:intial_hamiltonian}
\end{eqnarray}
Here $V$ is the interaction strength, $\sigma^k_x$ is a Pauli matrix and $n_k=\sigma^k_+\sigma^k_-$ is the Rydberg number operator with $\sigma^k_\pm$ being the raising and lowering operators of the $k$-th spin. This Hamiltonian is an Ising-model with long-range interactions in a transverse and a longitudinal field. Most experiments up to date use atomic states that interact via the van-der-Waals interaction, i.e. $\gamma=6$. We will focus here on this case but our approach also works for the dipole-dipole interacting case with $\gamma=3$. The Hamiltonian (\ref{eq:intial_hamiltonian}) was employed successfully to describe the excitation dynamics of driven Rydberg gases and has been proven to reflect very accurately the actual experimental situation \cite{Heidemann07,Heidemann08}.

The interaction between excited atoms strongly affects the excitation dynamics of the system through a mechanism which is called the Rydberg blockade \cite{Jaksch00}: The large interaction induced energy shift makes it virtually impossible to excite two nearby atoms simultaneously to the Rydberg state, i.e. the presence of one excited atom blocks the excitation of atoms in its vicinity (we assume $V\gg|\Delta|$ and $\Omega\gg V$ throughout). In what follows we will make this blockade effect manifest, which will create an effective three-body spin interaction in the Hamiltonian. Owed to the power law decay the strongest interaction takes place between nearest neighbors and we assume that a strict blockade is only present between them. We transform the Hamiltonian (\ref{eq:intial_hamiltonian}) into an interaction picture with respect to the nearest neighbor interaction by applying the unitary transformation
$U=\exp\left[ - i t V \sum_k^L n_k n_{k+1}\right]$. The first term of eq. (\ref{eq:intial_hamiltonian}) is the only one that does not commute with $U$ and one obtains
$U^\dagger \sigma_x^k U=\left[P_{k-1}+n_{k-1}e^{itV}\right]\sigma_+^k\left[P_{k+1}+n_{k+1}e^{itV}\right]+
\mathrm{h.c.}\nonumber$ where $P_k\equiv 1-n_k$. Note that both $n_k$ and $P_k$ are projection operators. Since $V\gg\Omega$ one can neglect the terms with rapidly oscillating phases which is essentially a rotating wave approximation. We then arrive at our working Hamiltonian
\begin{align}
  H=\Omega\sum_k^L P_{k-1}\sigma_x^k P_{k+1}+ \Delta \sum_k^L n_k + V\!\!\!\!\sum_{m>k+1}\frac{n_m n_k}{|k-m|^\gamma}\label{eq:RWA_hamiltonian}
\end{align}
where the first term is the blockade-induced three-body interaction: The excitation of an atom to a Rydberg state on site $k$ can only take place provided that both projectors $P_{k-1}$ and $P_{k+1}$ yield a non-zero value. This imposes a constraint such that the Hilbert space splits into uncoupled blocks each of which is characterized by the number of pairs of neighboring excited atoms. We will be concerned with the subspace in which there is no simultaneous excitation of nearest neighbors (referred to as physical subspace).

In the following we will show that for a certain set of parameters ($\Omega$, $\Delta$, $V$) Hamiltonian (\ref{eq:RWA_hamiltonian}) possesses an approximate analytical solution. The decisive idea is to add the term $H_\xi=\sum_k P_{k-1}P_{k+1}[\xi^{-1} n_k+\xi(1-n_k)]$ to Hamiltonian (\ref{eq:RWA_hamiltonian}) where $\xi$ is a real and positive parameter and subsequently subtract it. Obviously, adding and subtracting $H_\xi$ does not change $H$, but regrouping all terms conveniently allows us to rewrite the Hamiltonian as $H=E_0+H_\mathrm{3body}+H^\prime$ where now each term depends on $\xi$. Here $E_0=-\Omega L \xi$ will turn out to be the approximate ground state energy, $H_\mathrm{3body}$ is a spin Hamiltonian with three-body interactions that has an analytic ground state solution and $H^\prime$ is a perturbation. $H_\mathrm{3body}$ is given by
\begin{eqnarray}
  H_\mathrm{3body}=\Omega\sqrt{\xi^{-1}+\xi}\sum_k^L h_k=\Omega\sum_k^L h_k^\dagger h_k, \label{eq:RK_hamiltonian}
\end{eqnarray}
with
\begin{eqnarray}
  h_k=\sqrt{\frac{1}{\xi^{-1}+\xi}}P_{k-1}P_{k+1}\left[\sigma_x^k+\xi^{-1} n_k+\xi(1-n_k)\right].
\end{eqnarray}
The term of $h_k$ which contains $\sigma_x^k$ is proportional to the three-body interaction term in Hamiltonian (\ref{eq:RWA_hamiltonian}). The remaining ones are chosen such that $h_k$ becomes a self-adjoint operator with positive-semidefinite spectrum and only one non-zero eigenvalue: $\sqrt{\xi^{-1}+\xi}$. These are the terms that were introduced by $H_\xi$. The part that is canceling them is contained in the perturbation $H^\prime$ which we discuss later. At first, we construct the ground state $\left|\xi\right>$ of the Hamiltonian $H_\mathrm{3body}$. This state is annihilated by all $h_k$ and has hence energy zero. Note, that operators acting on neighboring sites do not commute, i.e. $\left[h_k,h_{k\pm1}\right]\neq 0$. It is thus not possible to use a local zero eigenstate of each $h_k$ and then form the total ground state by a product of these states. Instead one finds that the state of the physical subspace that is annihilated by all $h_k$ is given by
\begin{eqnarray}
   \left|\xi\right>=\frac{1}{\sqrt{Z_\xi}} \prod_k^L\left(1-\xi P_{k-1}\sigma_+^k P_{k+1}\right)\left|\downarrow\downarrow...\downarrow\right>. \label{eq:RK_state}
\end{eqnarray}
where $Z_\xi$ is a normalization constant. This state is a coherent superposition of all states that have no nearest neighbor excitations. The probability of each state is weighted by the factor $(\xi^2)^n$, where $n$ is the total number of Rydberg excitations in this state. This state is highly non-classical as it is a coherent superposition of all states from the physical subspace and cannot (except for $\xi=0$) be written as a product state. The existence of this ground state is due to the special projector property of each term of Hamiltonian (\ref{eq:RK_hamiltonian}) which is also known as Stochastic Matrix Form \cite{Castelnovo05}.

In order to calculate the normalization constant $Z_\xi$ one has to count the number of all allowed arrangement of excited atoms on the lattice and sum them using the weights $(\xi^2)^n$. Since there is strict nearest neighbor exclusion this sum is \emph{equivalent} to the partition function of a lattice gas of hard-core dimers, i.e. hard objects that occupy two neighboring lattice sites. In the limit $L\gg 1$ we obtain $Z_\xi=\left[(1/2)(1+\sqrt{1+4\xi^2})\right]^L$ such that we can identify $\xi^2$ as a fugacity. The fugacity suppresses/enhances the weight of a state with $n$ excited atoms or dimers by $(\xi^2)^n$ \cite{Goldenfield92}. We emphasize that the correspondence between the quantum problem and the dimer gas is solely formal. One striking difference is the range of the interaction: for the classical system only nearest neighbors interact while in the quantum system also interaction among next-nearest neighbors occur.

The aim is now to find a set of parameters ($\Omega$, $\Delta$, $V$) or a whole manifold of them such that $H^\prime$ is negligible compared to $H_\mathrm{3body}$. In this case the Hamiltonian of the Rydberg gas (\ref{eq:intial_hamiltonian}) is be very accurately approximated by $H_\mathrm{3body}$ for which we know the ground state. One finds that
\begin{eqnarray}
 H^\prime&=&\sum_k^L\left[\Delta+\Omega \left(3\xi-\xi^{-1}\right)+\left(2^{-\gamma}V-\Omega \xi\right)n_{k+2}\right]n_k\nonumber\\
  && + V\!\!\! \sum_{m>k+2} \frac{n_m n_k}{|k-m|^\gamma}\nonumber\\
  && -\Omega(\xi-\xi^{-1}) \sum_k n_k n_{k+1}(2-n_{k+2}).\label{eq:H_pert}
\end{eqnarray}
The first term of $H^\prime$ can be eliminated exactly provided that the conditions (i) $\Delta=-\Omega (3\xi-\xi^{-1})$ and (ii) $V=2^\gamma \Omega \xi$ are satisfied. The contribution of the second term in eq. (\ref{eq:H_pert}) is small since it accounts for the strongly diminished interaction between excited atoms that are at least three lattice sites apart. The third term vanishes exactly at $\xi=1$, but its contribution is negligible even away from this point since the probability for a simultaneous excitation of neighboring atoms is highly suppressed (even strictly zero in the physical subspace).

These considerations imply that upon meeting condition (ii), i.e. for an interaction strength satisfying $V=2^\gamma \Omega \xi$, the ground state energy of Hamiltonian (\ref{eq:RWA_hamiltonian}) is given by $E_0=-\xi \Omega L$ where $\xi=(1/6)[-(\Delta/\Omega)+\sqrt{12+(\Delta/\Omega)^2}]$. The latter relation is obtained directly from condition (i) and yields the conversion between the laser parameters and the square root of the fugacity, i.e. $\xi$.
\begin{figure}\center
\includegraphics[width=0.8\columnwidth]{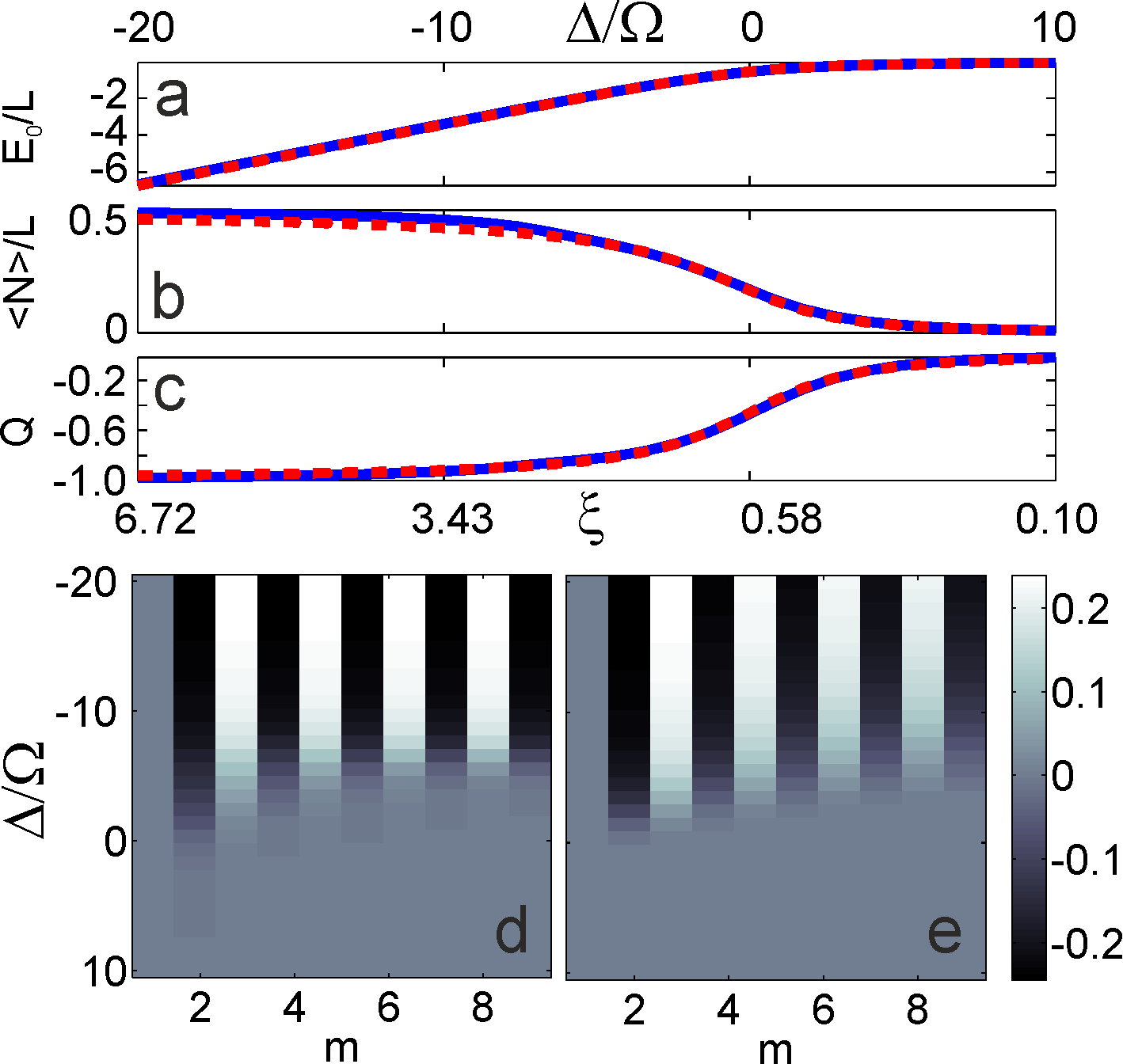}
\caption{Comparison between the numerical results obtained for a lattice with $L=20$ sites and $\gamma=6$ (blue) and the analytical expressions (dashed red). \textbf{a}: Energy per particle in the ground state, \textbf{b}: Mean density of Rydberg atoms on the lattice, \textbf{c}: Mandel Q-parameter of the Rydberg number distribution. \textbf{d}: Numerically calculated density-density correlation function. \textbf{e}: Density-density correlation function obtained for the state (\ref{eq:RK_state}). Note that at the same time as $\Delta$ also the potential is varied according to $V=2^\gamma \Omega \xi$. The values of $\xi$ are given underneath panel c.}\label{fig:EnQ}
\end{figure}
That this is indeed the case is shown in fig. \ref{fig:EnQ}a where we compare the ground state energy $E_0$ (red curve) with the numerical result (blue curve) obtained for a lattice with $L=20$ sites. The excellent agreement indicates that conditions (i) and (ii) define a manifold of approximate Rokhsar-Kivelson points \cite{Rokhsar88} in the parameter space ($\Omega$, $\Delta$, $V$) were the Hamiltonian of a gas of interacting Rydberg atoms (\ref{eq:intial_hamiltonian}) allows the approximate stochastic matrix form decomposition \cite{Castelnovo05} shown in eq. (\ref{eq:RK_hamiltonian}) and has the ground state (\ref{eq:RK_state}).

We can now calculate properties of the ground state of the system on this manifold in the same spirit in which we obtained the normalization constant $Z_\xi$. Expectation values of classical observables such as the mean number of excited atoms or density-density correlations then reduce to the manipulation of the partition function with fugacity $\xi^2$. The mean density of Rydberg atoms in the ground state is given by $\left<N\right>/L=\sum_k\left<\xi\right|n_k\left|\xi\right>/L=[1-1/(\sqrt{1+4 \xi^2})]/2$ which is shown in fig. \ref{fig:EnQ}b. We can furthermore obtain the full statistics of the Rydberg number distribution by taking derivatives of the partition function: The probability $p_k$ to count $k$ Rydberg atoms is given by $p_k=[(k!)^{-1}\partial^k_{\xi^2} Z_\xi\mid_{\xi=0}]/Z_\xi$. A common way for the experimental characterization of the distribution function is the Mandel Q-factor which quantifies the difference of the distribution $p_k$ from a Poissonian \cite{CubelLiebisch05}. This quantity, which is plotted in fig. \ref{fig:EnQ}c, evaluates to $Q=(\left<N^2\right>-\left<N\right>^2)/\left<N\right>-1=1/(2\sqrt{1+4\xi^2})-(1+8\xi^2)/(2+8\xi^2)$.
Except for $\xi=0$ it is negative showing a pronounced sub-Poissonian behavior which is expected for strongly interacting systems \cite{Ates06}.

A further important quantity characterizing the ground state is the connected density-density correlation function $g_{1,1+m}(\xi)=\left<n_1 n_{1+m}\right>-\left<n_1\right>\left<n_{1+m}\right>=\xi^2/(1+4\xi^2)[(\sqrt{1+4\xi^2}-2\xi^2-1)/(2\xi^2)]^m.$
It is shown in fig. \ref{fig:EnQ} in panels d and e together with the numerical result, both again in excellent agreement.
Visible correlations build up as soon as $\Delta<0$. They are exponentially decaying with the interparticle distance and alternating in sign, with anti-correlation between nearest neighbors. The corresponding correlation length is proportional to $\xi\,a$ and reaches the system size when $-\Delta/\Omega \approx 3L$.

\begin{figure}\center
\includegraphics[width=0.8\columnwidth]{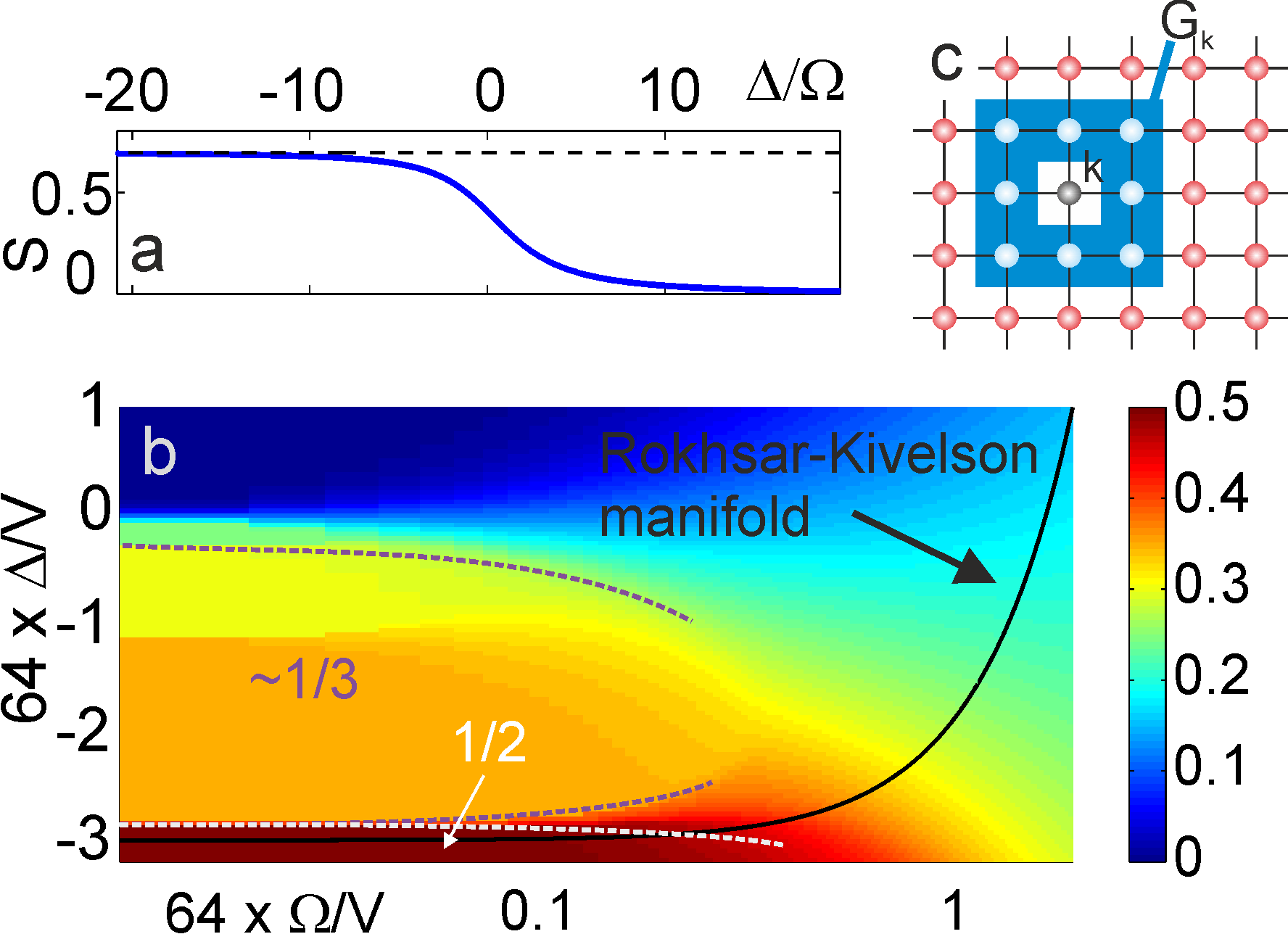}
\caption{\textbf{a}: Entanglement entropy $S$ of a single spin with the remaining ones. For large negative detuning the ground state becomes a GHZ state and $S$ reaches its maximum $\log 2$. \textbf{b}: Density of excited atoms as a function of the detuning and the Rabi frequency. The black line represents the set of parameters where the Rydberg gas ground state is approximately given by eq. (\ref{eq:RK_state}). Dashed lines are used as a guide to the eye delimiting the regions where the Rydberg density is approximately $1/3$ and $1/2$. \textbf{c}: The spin Hamiltonian (\ref{eq:RK_hamiltonian}) can be generalized to higher dimensions (here 2d) where the excitation on the $k$-th site (grey color) blocks the excitation of all sites contained in the set $G_k$.}\label{fig:phase_diag}
\end{figure}

We will now perform an analysis of the coherent properties of the system. To this end we study the reduced single particle density matrix $\rho_1(\xi)$ which allows us to quantify the entanglement of one spin with the rest of the system. We find
\begin{eqnarray*}
\rho_1(\xi)=(1/L)\left(
           \begin{array}{cc}
             \left<N\right> & -\left<N\right>/\xi \\
             -\left<N\right>/\xi & L-\left<N\right> \\
           \end{array}
         \right)
\end{eqnarray*}
which, except for $\xi=0$, represent a mixed state. This indicates entanglement of one atom with the remaining others which can be quantified by the entanglement entropy $S=-\mathrm{Tr} \rho_1(\xi)\, \log \rho_1(\xi)$. This function is shown in fig. \ref{fig:phase_diag}a. For large positive detuning, i.e. $\xi\approx 0$, the ground state is a product state $\left|\mathrm{init}\right>=\prod_k \left|g\right>_k$ and hence no entanglement is present. $S$ increases monotonously with $\xi$ and saturates at a value $\log 2$ for $\xi\rightarrow\infty$ which indicates maximal entanglement. Here the ground state is formally given by a GHZ state, which is the coherent superposition of the two possible anti-ferromagnetic states, i.e. $\left|\mathrm{GHZ}\right>=(1/\sqrt{2})[\left|\uparrow\downarrow\uparrow\downarrow...\right>+\left|\downarrow\uparrow\downarrow\uparrow...\right>]$ (even number of sites assumed).

The above considerations indicate that the typical experimental initial state $\left|\mathrm{init}\right>$ (no Rydberg atoms present) can be adiabatically connected to the fully entangled GHZ state by varying $\xi$ from zero to infinity, i.e. by varying $\Omega$ and $\Delta$ in time. Experimentally this is usually done at fixed interaction strength $V$. The approximate manifold of Rokhsar-Kivelson points is then given through $(2^\gamma\Omega/V)^2-(2^\gamma\Delta/V)=3$ which is obtained from (i) and (ii) and shown as the black curve in fig. \ref{fig:phase_diag}b. The GHZ state is obtained by initially choosing a large positive detuning and following this curve until one reaches $\Delta_\mathrm{min}=-3/2^\gamma\,V$, i.e. $\Omega_\mathrm{min}=0$. Performing this process adiabatically becomes increasingly difficult as the number of particles increases due to an ever closing energy gap. Eventually, this will lead to symmetry breaking which singles out one of the two anti-ferromagnetic states or leads to domain formation. Experiments have to be carried out on a time shorter than the lifetime of the atomic Rydberg state (typically $100\,\mu\mathrm{s}$ for Rubidium and a principal quantum number in the range $n=40...70$). It is indeed possible to find experimental parameters that achieve that (see Refs. \cite{Pohl10,Schachenmayer10,Olmos10-1,Lesanovsky10}).

Let us finally discuss the generalization of Hamiltonian (\ref{eq:RK_hamiltonian}) to higher dimensions and blockade ranges that can go beyond the nearest neighbors. To this end we replace the product $P_{k-1}P_{k+1}$ by an operator which projects onto the state $\prod_{q\epsilon G_k}\left|\downarrow\right>_q$. Here $G_k$ is a set that contains the indices of lattice sites that surround the $k$-th site, i.e. that are blocked when spin $k$ is excited (see fig. \ref{fig:phase_diag}c). The ground state of this Hamiltonian is then constructed analogous to the state (\ref{eq:RK_state}) with the constraint being that a simultaneous excitation on site $k$ and on any of the sites contained in $G_k$ is forbidden. Calculations of expectation values here again reduce to the manipulation of a partition sum of a classical system of hard objects. It is not immediately evident whether such models actually represent an experimentally relevant system. This depends on whether conditions similar to (i) and (ii) can be found which cancel the unwanted many-body terms in $H^\prime$. However, the knowledge of the ground state is valuable, e.g. for performing perturbation theory in order to move away from the exactly solvable situation.

Funding through EPSRC and fruitful discussions with J.P. Garrahan, G. Adesso, B. Olmos, P. Kr\"uger and M. M\"uller are gratefully acknowledged.


\begin{thebibliography}{23}
\expandafter\ifx\csname natexlab\endcsname\relax\def\natexlab#1{#1}\fi
\expandafter\ifx\csname bibnamefont\endcsname\relax
  \def\bibnamefont#1{#1}\fi
\expandafter\ifx\csname bibfnamefont\endcsname\relax
  \def\bibfnamefont#1{#1}\fi
\expandafter\ifx\csname citenamefont\endcsname\relax
  \def\citenamefont#1{#1}\fi
\expandafter\ifx\csname url\endcsname\relax
  \def\url#1{\texttt{#1}}\fi
\expandafter\ifx\csname urlprefix\endcsname\relax\def\urlprefix{URL }\fi
\providecommand{\bibinfo}[2]{#2}
\providecommand{\eprint}[2][]{\url{#2}}

\bibitem[{\citenamefont{Sachdev}(1999)}]{Sachdev99}
\bibinfo{author}{\bibfnamefont{S.}~\bibnamefont{Sachdev}},
  \emph{\bibinfo{title}{Quantum Phase Transitions}}
  (\bibinfo{publisher}{Cambridge University Press, Cambridge, UK},
  \bibinfo{year}{1999}).

\bibitem[{\citenamefont{Affleck et~al.}(1988)\citenamefont{Affleck, Kennedy,
  Lieb, and Tasaki}}]{Affleck88}
\bibinfo{author}{\bibfnamefont{I.}~\bibnamefont{Affleck}} \textit{et al.},
  \bibinfo{journal}{Comm. Math. Phys.} \textbf{\bibinfo{volume}{115}},
  \bibinfo{pages}{477} (\bibinfo{year}{1988}).

\bibitem[{\citenamefont{Bloch et~al.}(2008)\citenamefont{Bloch, Dalibard, and
  Zwerger}}]{Bloch08}
\bibinfo{author}{\bibfnamefont{I.}~\bibnamefont{Bloch}},
  \bibinfo{author}{\bibfnamefont{J.}~\bibnamefont{Dalibard}}, \bibnamefont{and}
  \bibinfo{author}{\bibfnamefont{W.}~\bibnamefont{Zwerger}},
  \bibinfo{journal}{Rev. Mod. Phys.} \textbf{\bibinfo{volume}{80}},
  \bibinfo{eid}{885} (\bibinfo{year}{2008}).

\bibitem[{\citenamefont{Gallagher}(1984)}]{Gallagher84}
\bibinfo{author}{\bibfnamefont{T.}~\bibnamefont{Gallagher}},
  \emph{\bibinfo{title}{Rydberg Atoms}} (\bibinfo{publisher}{Cambridge
  University Press}, \bibinfo{year}{1984}).

\bibitem[{\citenamefont{Saffman et~al.}(2010)\citenamefont{Saffman, Walker, and
  M\o{}lmer}}]{Saffman10-2}
\bibinfo{author}{\bibfnamefont{M.}~\bibnamefont{Saffman}},
  \bibinfo{author}{\bibfnamefont{T.~G.} \bibnamefont{Walker}},
  \bibnamefont{and}
  \bibinfo{author}{\bibfnamefont{K.}~\bibnamefont{M\o{}lmer}},
  \bibinfo{journal}{Rev. Mod. Phys.} \textbf{\bibinfo{volume}{82}},
  \bibinfo{pages}{2313} (\bibinfo{year}{2010}).

\bibitem[{\citenamefont{Heidemann et~al.}(2007)\citenamefont{Heidemann,
  Raitzsch, Bendkowsky, Butscher, L\"{o}w, Santos, and Pfau}}]{Heidemann07}
\bibinfo{author}{\bibfnamefont{R.}~\bibnamefont{Heidemann}} \textit{et al.},
  \bibinfo{journal}{Phys. Rev. Lett.} \textbf{\bibinfo{volume}{99}},
  \bibinfo{eid}{163601} (\bibinfo{year}{2007}).

\bibitem[{\citenamefont{Reetz-Lamour et~al.}(2008)\citenamefont{Reetz-Lamour,
  Amthor, Deiglmayr, and Weidem\"{u}ller}}]{Reetz-Lamour08}
\bibinfo{author}{\bibfnamefont{M.}~\bibnamefont{Reetz-Lamour}} \textit{et al.},
  \bibinfo{journal}{Phys. Rev. Lett.} \textbf{\bibinfo{volume}{100}},
  \bibinfo{eid}{253001} (\bibinfo{year}{2008}).

\bibitem[{\citenamefont{Urban et~al.}(2009)\citenamefont{Urban, Johnson,
  Henage, Isenhower, Yavuz, Walker, and Saffman}}]{Urban08}
\bibinfo{author}{\bibfnamefont{E.}~\bibnamefont{Urban}} \textit{et al.},
  \bibinfo{journal}{Nature Phys.} \textbf{\bibinfo{volume}{5}},
  \bibinfo{pages}{110} (\bibinfo{year}{2009}).

\bibitem[{\citenamefont{Ga\"{e}tan et~al.}(2009)\citenamefont{Ga\"{e}tan,
  Miroshnychenko, Wilk, Chotia, Viteau, Comparat, Pillet, Browaeys, and
  Grangier}}]{Gaetan08}
\bibinfo{author}{\bibfnamefont{A.}~\bibnamefont{Ga\"{e}tan}} \textit{et al.},
  \bibinfo{journal}{Nature Phys.} \textbf{\bibinfo{volume}{5}},
  \bibinfo{pages}{115} (\bibinfo{year}{2009}).

\bibitem[{\citenamefont{Olmos et~al.}(2009)\citenamefont{Olmos,
  Gonz\'{a}lez-F\'{e}rez, and Lesanovsky}}]{Olmos09-3}
\bibinfo{author}{\bibfnamefont{B.}~\bibnamefont{Olmos}},
  \bibinfo{author}{\bibfnamefont{R.}~\bibnamefont{Gonz\'{a}lez-F\'{e}rez}},
  \bibnamefont{and}
  \bibinfo{author}{\bibfnamefont{I.}~\bibnamefont{Lesanovsky}},
  \bibinfo{journal}{Phys. Rev. Lett.} \textbf{\bibinfo{volume}{103}},
  \bibinfo{pages}{185302} (\bibinfo{year}{2009}).

\bibitem[{\citenamefont{Weimer et~al.}(2008)\citenamefont{Weimer, L\"{o}w,
  Pfau, and B\"{u}chler}}]{Weimer08}
\bibinfo{author}{\bibfnamefont{H.}~\bibnamefont{Weimer}} \textit{et al.},
  \bibinfo{journal}{Phys. Rev. Lett.} \textbf{\bibinfo{volume}{101}},
  \bibinfo{pages}{250601} (\bibinfo{year}{2008}).

\bibitem[{\citenamefont{Lesanovsky et~al.}(2010)\citenamefont{Lesanovsky,
  Olmos, and Garrahan}}]{Lesanovsky10}
\bibinfo{author}{\bibfnamefont{I.}~\bibnamefont{Lesanovsky}},
  \bibinfo{author}{\bibfnamefont{B.}~\bibnamefont{Olmos}}, \bibnamefont{and}
  \bibinfo{author}{\bibfnamefont{J.~P.} \bibnamefont{Garrahan}},
  \bibinfo{journal}{Phys. Rev. Lett.} \textbf{\bibinfo{volume}{105}},
  \bibinfo{pages}{100603} (\bibinfo{year}{2010}).

\bibitem[{\citenamefont{Pohl et~al.}(2010)\citenamefont{Pohl, Demler, and
  Lukin}}]{Pohl10}
\bibinfo{author}{\bibfnamefont{T.}~\bibnamefont{Pohl}},
  \bibinfo{author}{\bibfnamefont{E.}~\bibnamefont{Demler}}, \bibnamefont{and}
  \bibinfo{author}{\bibfnamefont{M.~D.} \bibnamefont{Lukin}},
  \bibinfo{journal}{Phys. Rev. Lett.} \textbf{\bibinfo{volume}{104}},
  \bibinfo{pages}{043002} (\bibinfo{year}{2010}).

\bibitem[{\citenamefont{Schachenmayer et~al.}(2010)\citenamefont{Schachenmayer,
  Lesanovsky, and Daley}}]{Schachenmayer10}
\bibinfo{author}{\bibfnamefont{J.}~\bibnamefont{Schachenmayer}},
  \bibinfo{author}{\bibfnamefont{I.}~\bibnamefont{Lesanovsky}},
  \bibnamefont{and} \bibinfo{author}{\bibfnamefont{A.}~\bibnamefont{Daley}},
  \bibinfo{journal}{New J. Phys.}
  \textbf{\bibinfo{volume}{12}}
  \bibinfo{pages}{103044}
  (\bibinfo{year}{2010}).

\bibitem[{\citenamefont{Weimer and B\"uchler}(2010)}]{Weimer10-2}
\bibinfo{author}{\bibfnamefont{H.}~\bibnamefont{Weimer}} \bibnamefont{and}
  \bibinfo{author}{\bibfnamefont{H.~P.} \bibnamefont{B\"uchler}},
  \bibinfo{journal}{preprint} p. \bibinfo{pages}{arXiv:1007.2189}
  (\bibinfo{year}{2010}).

\bibitem[{\citenamefont{Rokhsar and Kivelson}(1988)}]{Rokhsar88}
\bibinfo{author}{\bibfnamefont{D.~S.} \bibnamefont{Rokhsar}} \bibnamefont{and}
  \bibinfo{author}{\bibfnamefont{S.~A.} \bibnamefont{Kivelson}},
  \bibinfo{journal}{Phys. Rev. Lett.} \textbf{\bibinfo{volume}{61}},
  \bibinfo{pages}{2376} (\bibinfo{year}{1988}).

\bibitem[{\citenamefont{Castelnovo et~al.}(2005)\citenamefont{Castelnovo,
  Chamon, Mudry, and Pujol}}]{Castelnovo05}
\bibinfo{author}{\bibfnamefont{C.}~\bibnamefont{Castelnovo}} \textit{et al.},
  \bibinfo{journal}{Annals of Physics} \textbf{\bibinfo{volume}{318}},
  \bibinfo{pages}{316} (\bibinfo{year}{2005}).

\bibitem[{\citenamefont{Heidemann et~al.}(2008)\citenamefont{Heidemann,
  Raitzsch, Bendkowsky, Butscher, L\"{o}w, and Pfau}}]{Heidemann08}
\bibinfo{author}{\bibfnamefont{R.}~\bibnamefont{Heidemann}} \textit{et al.},
  \bibinfo{journal}{Phys. Rev. Lett.} \textbf{\bibinfo{volume}{100}},
  \bibinfo{eid}{033601} (\bibinfo{year}{2008}).

\bibitem[{\citenamefont{Jaksch et~al.}(2000)\citenamefont{Jaksch, Cirac,
  Zoller, Rolston, C\^{o}t\'{e}, and Lukin}}]{Jaksch00}
\bibinfo{author}{\bibfnamefont{D.}~\bibnamefont{Jaksch}} \textit{et al.},
  \bibinfo{journal}{Phys. Rev. Lett.} \textbf{\bibinfo{volume}{85}},
  \bibinfo{pages}{2208} (\bibinfo{year}{2000}).

\bibitem[{\citenamefont{Goldenfield}(1992)}]{Goldenfield92}
\bibinfo{author}{\bibfnamefont{N.}~\bibnamefont{Goldenfield}},
  \emph{\bibinfo{title}{Lectures on Phase Transitions and the Renormalization
  Group}} (\bibinfo{publisher}{Westview Press}, \bibinfo{year}{1992}).

\bibitem[{\citenamefont{Liebisch et~al.}(2005)\citenamefont{Liebisch, Reinhard,
  Berman, and Raithel}}]{CubelLiebisch05}
\bibinfo{author}{\bibfnamefont{T.~C.} \bibnamefont{Liebisch}} \textit{et al.},
  \bibinfo{journal}{Phys. Rev. Lett.} \textbf{\bibinfo{volume}{95}},
  \bibinfo{pages}{253002} (\bibinfo{year}{2005}).

\bibitem[{\citenamefont{Ates et~al.}(2006)\citenamefont{Ates, Pohl, Pattard,
  and Rost}}]{Ates06}
\bibinfo{author}{\bibfnamefont{C.}~\bibnamefont{Ates}} \textit{et al.},
  \bibinfo{journal}{Journal of Physics B} \textbf{\bibinfo{volume}{39}},
  \bibinfo{pages}{L233} (\bibinfo{year}{2006}).

\bibitem[{\citenamefont{Olmos et~al.}(2010)\citenamefont{Olmos,
  Gonz\'alez-F\'erez, and Lesanovsky}}]{Olmos10-1}
\bibinfo{author}{\bibfnamefont{B.}~\bibnamefont{Olmos}},
  \bibinfo{author}{\bibfnamefont{R.}~\bibnamefont{Gonz\'alez-F\'erez}},
  \bibnamefont{and}
  \bibinfo{author}{\bibfnamefont{I.}~\bibnamefont{Lesanovsky}},
  \bibinfo{journal}{Phys. Rev. A} \textbf{\bibinfo{volume}{81}},
  \bibinfo{pages}{023604} (\bibinfo{year}{2010}).

\end{thebibliography}
\end{document}